\documentclass[10pt,aps,prb,twocolumn,amsmath,amssymb]{revtex4}
\usepackage{amsmath}
\usepackage{graphicx}
\usepackage{color}
\begin{document}
 \title{Near exact excited states of the carbon dimer in a quadruple-zeta basis using
a general non-Abelian density matrix renormalization group algorithm}
\author{Sandeep Sharma}
\email{sanshar@gmail.com}
\affiliation{Department of Chemistry, Frick Laboratory, Princeton University, NJ 08544}
\begin{abstract}
 We extend our previous work [J. Chem. Phys, \textbf{136}, 124121], which described a spin-adapted (SU(2) symmetry) density matrix renormalization group (DMRG) algorithm, to additionally utilize
general non-Abelian point group symmetries. A key strength of the present formulation is that the requisite tensor operators are
not hard-coded for each symmetry group, but are instead generated on the fly 
using the appropriate Clebsch-Gordan coefficients. This allows our
single implementation to easily enable (or disable) any non-Abelian point group symmetry (including SU(2) spin symmetry). 
We use our implementation to compute the ground state potential energy curve 
of the C$_2$ dimer in the cc-pVQZ basis set (with a frozen-core), corresponding to a Hilbert space dimension of 10$^{12}$ many-body 
states. While our calculated energy lies within the 0.3 mE$_h$ error bound of previous initiator full configuration interaction quantum Monte Carlo (i-FCIQMC) and correlation energy extrapolation by intrinsic scaling (CEEIS) calculations, our estimated residual error is
only 0.01 mE$_h$, much more accurate than these previous estimates. Further, due to the additional efficiency afforded by the algorithm, we compute the potential energy curves of 
twelve states: the three lowest  levels for each of the irreducible representations $^1\Sigma^+_g$,  $^1\Sigma^+_u$, $^1\Sigma^-_g$ and $^1\Sigma^-_u$,  to an estimated accuracy within 0.1 mE$_h$ of the exact result in this basis.
 \end{abstract}
\maketitle

The density matrix renormalization group (DMRG) was first introduced  over 20 years ago\cite{White1992, White1993} and has since become an indispensable part of the toolbox of chemists and 
physicists. Its application in quantum chemistry began with the seminal paper by  White and Martin\cite{whiteqm}, and was soon followed with work by many other groups\cite{chan2011,legeza-rev,marti11,yuki2,wouters-review,Kurashige2013, sharma2014}. Recently, the DMRG has seen many important algorithmic advancements such as orbital optimization\cite{Zgid2008dmrgscf, Ghosh2008}, combination with perturbation theory\cite{yuki,dmrgpt22}, configuration interaction\cite{dmrgmrci}, and canonical transformation theory\cite{dmrgct}, and has also been incorporated into several major quantum chemistry packages 
such as \textsc{Molpro}\cite{werner2012}, \textsc{\mbox{Q-chem}}\cite{qchem14}  and \textsc{Orca}\cite{orca12}. 

Significant improvements in DMRG efficiency can be gained by utilizing non-Abelian symmetries such as SU(2) symmetry.
This was first attempted by Nishino using interaction-round-a-face DMRG (IRF-DMRG)\cite{sierra}. This was later followed by McCulloch's work\cite{mcculloch1,mcculloch3} that introduced
the Wigner-Eckart theorem to gain large efficiencies and the ``quasi-density matrix" to ensure that only spin multiplet states are retained during renormalization. Zgid and Nooijen\cite{Zgid2008spin} later implemented an algorithm for targeting specific spin states with general quantum chemistry Hamiltonians using the quasi-density matrix, but did not take advantage of the Wigner-Eckart theorem,
and thus did not provide a boost in efficiency. The first efficient spin adapted code for quantum chemistry was presented independently by Wouters et al\cite{wouters} and Sharma and Chan\cite{sharma2012spin}. Weichselbaum\cite{Weichselbaum2012} then presented a general framework for using non-Abelian symmetries in tensor networks and demonstrated how the so-called ``inner'' and ``outer'' multiplicities can be appropriately treated. Here, we present an algorithm for utilizing the most general non-Abelian symmetry found in a molecule: SU(2) $\times$  non-Abelian point group symmetry, in conjunction with a quantum chemical Hamiltonian, where the added complication of treating outer multiplicities is not present. This extends our 
previous work\cite{sharma2012spin}, which we refer to for a detailed discussion of the algorithm and notation. We use our new algorithm  to calculate frozen core near-exact potential
curves for the 
ground and several excited states of the C$_2$ molecule at various bond lengths in a cc-pVQZ basis set\cite{dunningbasis}, to 10 $\mu$E$_h$ and 0.1 mE$_h$ accuracies, respectively. 

The sweep algorithm, introduced by White, is commonly used to optimize the underlying matrix product state (MPS) wavefunction of DMRG. The algorithm can be sub-divided into 
three steps: blocking, wavefunction solution and decimation. In the following we take each step of the algorithm and show how it is modified to use non-Abelian symmetry. However, before this, 
it is appropriate to reiterate that the central concept on which the algorithm hinges is the Wigner-Eckart theorem, shown in Equation~\ref{eq:we}, where $\psi^{\gamma_i}$ and $\phi^{\beta_k}$ are states belonging to the $i$ and $k$ rows of irreducible representations $\gamma$ and $\beta$ respectively, and $\hat{O}^{\alpha_j}$ is the tensor operator of irreducible representation $\alpha$ that transforms as its $j$th row.
\begin{align}
 \langle \psi^{\gamma_i} | \hat{O}^{\alpha_j} | \phi^{\beta_k}\rangle = \langle\beta_k\alpha_j |\gamma_i\rangle \langle\psi^\gamma||\hat{O}^\alpha||\phi^\beta\rangle \label{eq:we}
\end{align}
This equation states, in essence, that $n_\gamma n_\beta$ matrix elements of $n_\alpha $ different operators can be stored using just one ``reduced matrix element'' ($\langle\psi^\gamma||\hat{O}^\alpha||\phi^\beta\rangle$),  as long as the Clebsch-Gordan coefficients $\langle\beta_k \alpha_j |\gamma_i\rangle$ of the symmetry group are known. To use the Wigner-Eckart theorem, the design of the algorithm should ensure that only those reduced sets of operators and states that transform as irreducible representation of a given non-abelian group explicitly appear.
\vspace{2 mm}

\noindent\textit{Blocking}: In this step  a single site ($\bullet$) is added to the left block ($\mathcal{L}$) (containing $k$ sites) to generate a resulting system block ($\mathcal{S}$) containing $k+1$ sites. The system block in general will contain $O(k^2)$ normal (or complementary) operators, each of which can be multiplied with one of the appropriate $O(k^2)$ complementary (or normal operators) on the environment block to form the Hamiltonian (although we note that an explicit matrix representation of this Hamiltonian is never constructed). The definition of these normal operators and corresponding complementary operators with which they are multiplied are given in the two columns of Table~\ref{tab:ops}.

\begin{table}
\caption{Normal and complementary operators used in the usual DMRG algorithm, where the indices are spin indices. In the definition of complementary operators the repeated indices are summed over.} \label{tab:ops}
 \begin{tabular}{lccl}
 \hline
 \hline
 Normal&&& Complementary \\
Operator&&& Operator\\
 \cline{1-1}\cline{4-4}
 $ \hat{I}$ &&&$\hat{H}=t_{ij}a_i^\dag a_j + \langle ij|kl\rangle a_i^\dag a_j^\dag a_l a_k$ \\
 $a_i^\dag$ &&& $\hat{R}_i = t_{ij} a_j + \langle ij|kl\rangle a_j^\dag a_l a_k$\\
 $\hat{A}_{ij} = a_i^\dag a_j^\dag$ &&& $\hat{P}_{ij} =  \langle ij|kl\rangle  a_l a_k$\\
 $\hat{B}_{ij} = a_i^\dag a_j$ &&& $\hat{Q}_{ij} =  (\langle ik|jl\rangle -\langle ik|lj\rangle) a_k^\dag a_l$\\
 \hline
  \end{tabular}
\end{table}


Unfortunately, the usual normal and complementary operators cannot be used directly with non-Abelian symmetries because they in general do not transform as irreducible 
representations of the group. To work with tensor operators, we start with some notation. First, we denote the creation/destruction tensor operator $a_I^{\gamma\dag}$/$a_I^{\gamma}$ of a ``spatial orbital'' $I$ as a set of $n_\gamma$ creation/destruction operators $\{a_I^{\gamma_i\dag}\}$/$\{p_{\gamma_i}a_I^{\gamma_i}\}$, one for each ``spin orbital'' $i$ of the spatial orbital $I$. Here, $p_{\gamma_i}$ is a phase factor which in general is not 1 and depends on the phase convention used in the construction of Clebsch-Gordan coefficients. It ensures that the destruction operators also transform as a set of tensor operators of irreducible representation $\gamma$\cite{brink,sharma2012spin}. (A Greek letter is used to denote the irreducible representation and the capitalized Roman letter denotes the spatial orbital.) Note, here the label ``spin'' and ``spatial'' is used more generally than in the context of spin symmetry alone. A spatial orbital is the set of all orbitals that transform as different rows of an irreducible representation. For example, when one is using the SU(2) group then each spatial orbital has only 2 spin orbitals, however, when one is using the SU(2) $\times D_{\infty h}$ group, then a spatial orbital belonging to irrep $\Pi_g$ will have 4 spin orbitals.

When two creation tensor operators $a_I^{\alpha\dag}$ and $a_J^{\beta\dag}$ are multiplied, they form a new set of $n_\alpha n_\beta$ operators that do not themselves form a set of tensor operators, but which can be converted into multiplets of tensor operators using the Clebsch-Gordan coefficients as shown in Equation~\ref{eq:topcg}. 
\begin{align}
CC_{IJ}^{\gamma_k} =& \sum_{ij} CC_{IJ}^{\gamma_k[\alpha_i\beta_j]} =\sum_{ij} \langle \alpha_i \beta_j|\gamma_k\rangle a_I^{\alpha_i\dag}a_J^{\beta_j\dag} \label{eq:topcg}\\
CCD_{IJK} ^{\delta_l} =& \sum_{ijkm} CCD_{IJK}^{\delta_l[\alpha_i \theta_m[\beta_j\gamma_k]]}\nonumber\\
=&\sum_{ijkm} \langle \beta_j\gamma_k|\theta_m\rangle \langle\alpha_i\theta_m|\delta_l\rangle a_I^{\alpha_i\dag}a_J^{\beta_j\dag}a_K^{\gamma_k} \label{eq:topcg2}
\end{align}
Here the intermediate $CC_{IJ}^{\gamma_k[\alpha_i\beta_j]}=\langle \alpha_i \beta_j|\gamma_k\rangle a_I^{\alpha_i\dag}a_J^{\beta_j\dag}$ is defined for convenience. When more than two tensor operators are multiplied (see Equation~\ref{eq:topcg2}), then to uniquely define the final tensor operator ($CCD_{IJK} ^{\delta_l}$) the irreducible representation of the intermediate tensor operator ($\theta$) and the order in which the operators are multiplied ($\delta_l[\alpha_i \theta_m[\beta_j\gamma_k]]$) need to be specified as well.

 As a short-hand for the product of two tensor operators of the form specified in Equation~\ref{eq:topcg} we define the symbol $\otimes^\gamma$, which generates $n_\gamma$ tensor operators $\hat{N}^\gamma$ by taking an outer product of two tensor operators $\hat{L}^\alpha$ and  $\hat{M}^\beta$ by using the formula given in Equation~\ref{eq:defop}, where the same notation as the one used for the intermediate ($CC_{IJ}^{\gamma_k[\alpha_i\beta_j]}$) from Equation~\ref{eq:topcg} is used on the RHS of the definition. 
 \begin{align}
\mathrm{Notation~~~~~~~~~~~~~~} & \mathrm{Definition}\nonumber\\
\hat{N}^{\gamma[\alpha\beta]} = \hat{L}^\alpha\otimes^\gamma\hat{M}^\beta \mathrm{~~}&   \{\hat{N}^{\gamma_k} = \sum_{ij} \hat{N}^{\gamma_k [\alpha_i\beta_j]} \}\label{eq:defop}
\end{align}
With these notations in place, we can conveniently define a new set of normal and complementary operators as shown in Table~\ref{tab:tops}, to replace the operators in Table~\ref{tab:ops}.
\begin{align}
& \mathbf{H}[\mathcal{S}] = \mathbf{H}[\mathcal{L}] \otimes^{A_g} \mathbf{1}[\bullet] + \mathbf{H}[\bullet] \otimes^{A_g} \mathbf{1}[\mathcal{L}] \nonumber\\
 &+  (1/2)  \sum_{I\in \mathcal{L}} \left(\mathbf{a}^{\gamma\dag}_J[\mathcal{L}] \otimes^{A_g} \mathbf{R}^{\gamma}_{J}[\bullet]  + \mathbf{a}^{\gamma}_J[\mathcal{L}] \otimes^{A_g} \mathbf{R}^{\gamma\dag}_{J}[\bullet]\right) \nonumber\\
&+(1/2)  \sum_{I\in \bullet} \left(\mathbf{a}^{\gamma\dag}_J[\bullet] \otimes^{A_g} \mathbf{R}^{\gamma}_{J}[\mathcal{L}]  + \mathbf{a}^{\gamma}_J[\bullet] \otimes^{A_g} \mathbf{R}^{\gamma\dag}_{J}[\mathcal{L}]\right) \nonumber\\
&+ (1/2)  \sum_{IJ\in \bullet} \left(\mathbf{A}^{\gamma}_{IJ}[\bullet] \otimes^{A_g} \mathbf{P}^{\gamma}_{IJ}[\mathcal{L}] + \mathbf{A}^{\gamma\dag}_{IJ}[\bullet] \otimes^{A_g} \mathbf{P}^{\gamma\dag}_{IJ}[\mathcal{L}] \right)\nonumber\\
&+(1/2)  \sum_{IJ\in \bullet} \left(\mathbf{B}^{\gamma}_{IJ}[\bullet] \otimes^{A_g} \mathbf{Q}^{\gamma}_{IJ}[\mathcal{L}] \right)\nonumber\\
 &\mathbf{R}_I^\alpha[\mathcal{S}] = \mathbf{R}_I^\alpha[\mathcal{L}] \otimes^\alpha \mathbf{1}[\bullet] + \mathbf{R}_I^\alpha[\bullet] \otimes^\alpha \mathbf{1}[\mathcal{L}] \nonumber\\
& + \sum_{J\in \bullet} W(\alpha\beta A_g\gamma; \gamma \alpha) \left(2\mathbf{P}^{\gamma}_{IJ}[\mathcal{L}]\otimes^\alpha \mathbf{a}^{\beta\dag}_J[\bullet] + \mathbf{Q}^{\gamma}_{IJ}[\mathcal{L}]\otimes^\alpha \mathbf{a}^{\beta}_{J}[\bullet]\right) \nonumber\\
 &+ \sum_{J\in \mathcal{L}} W(\alpha\beta A_g\gamma; \gamma \alpha) \left(2\mathbf{P}^{\gamma}_{IJ}[\bullet]\otimes^\alpha \mathbf{a}^{\beta}_J[\mathcal{L}] + \mathbf{Q}^{\gamma}_{IJ}[\bullet]\otimes^\alpha \mathbf{a}^{\beta}_J[\mathcal{L}]\right) \nonumber\\
&\mathbf{P}^\gamma_{IJ}[\mathcal{S}] = \mathbf{P}^\gamma_{IJ}[\mathcal{L}] \otimes^\gamma \mathbf{1}[\bullet] + \mathbf{P}^\gamma_{\IJ}[\bullet] \otimes^\gamma \mathbf{1}[\mathcal{L}] \nonumber\\
&+\sum_{\substack{ K\in \bullet \\L \in\mathcal{L}} } \left(\sum_{\delta_k\xi_l\alpha_i\beta_j}\frac{\nu^{\alpha_i\beta_j\delta_k\xi_l}_{IJKL}\langle-\gamma_m|\xi_l\delta_k\rangle\langle\gamma_m|\alpha_i\beta_j\rangle}{\langle\gamma_m -\gamma_m|Ag\rangle}\right) \mathbf{a}^\xi_L[\mathcal{L}] \otimes^\gamma \mathbf{a}^{\delta}_K[\bullet] \nonumber\\
&\mathbf{Q}^\gamma_{IJ}[\mathcal{S}] = \mathbf{Q}^\gamma_{IJ}[\mathcal{L}] \otimes^\gamma \mathbf{1}[\bullet] + \mathbf{Q}^\gamma_{\IJ}[\bullet] \otimes^\gamma \mathbf{1}[\mathcal{L}]+ \nonumber\\
&\sum_{\substack{ K\in \bullet \\L \in\mathcal{L}} } \left(\sum_{\delta_k\xi_l\alpha_i\beta_j}\frac{(\nu^{\alpha_i\delta_k\beta_j\xi_l}_{IKJL}-\nu^{\alpha_i\delta_k\xi_l\beta_j}_{IKLJ})\langle-\gamma_m|\xi_l\delta_k\rangle\langle\gamma_m|\alpha_i\beta_j\rangle}{\langle\gamma_m -\gamma_m|Ag\rangle} \right)\nonumber\\
&~~~~~~~~~ \mathbf{a}^{\delta\dag}_K[\bullet] \otimes^\gamma \mathbf{a}^\xi_L[\mathcal{L}] \label{eq:topgen}
\end{align}

These operators are constructed during the blocking step using the formulae in Equation~\ref{eq:topgen} (only formulae for the construction of complementary operators are shown), where the quantity $W(\alpha\beta A_g\gamma; \gamma \alpha)$ is the Racah W-coefficient\cite{brink}. In these formulae, the operators on the left ($\mathcal{L}$) and dot ($\bullet$) blocks are multiplied using the outer product notation given in Equation~\ref{eq:defop}.


\begin{table*}
\caption{Definitions of the normal and complementary operators used in the non-Abelian symmetry adapted DMRG algorithm.  $C$ and $D$ denote the creation and destruction operators, for example ``$CDD_{IJK}$'' represents the product of three tensor operators, one creation on spatial orbital I and two destructions on spatial orbitals J and K. The superscript $A_g$ is used to denote the fully symmetric irreducible representation of the symmetry group of interest. The symmetry label with a negative sign (e.g. $-\alpha_i$) before it, represents the row of the irreducible representation ($\alpha$) that can couple to the original irreducible representation ($\alpha_i$) to give a fully symmetric irreducible 
representation ($A_g$), i.e. $\langle\alpha_i -\alpha_i|Ag\rangle \neq 0$.} \label{tab:tops}
 \begin{tabular}{lccl}
 \hline
 \hline
 Normal&&& Complementary \\
Operator&&& Operator\\
 \cline{1-1}\cline{4-4}
 $ \hat{I}$ &&&$\hat{H}=\sum_{\alpha_i\beta_j}t^{\alpha_i \beta_j}_{IJ}a^{\alpha_i\dag}_I a^{\beta_j}_J + \sum\nu^{\alpha_i\beta_j\delta_k\xi_l}_{IJKL}a^{\alpha_i\dag}_I a^{\beta_j\dag}_J a^{\xi_l}_L a^{\delta_k}_K $ \\
 $a_I^{\alpha_i\dag}$ &&& $\hat{R}_I^{\alpha_i} = \sum_{IJ\alpha_I} t_{IJ}^{\alpha_i -\alpha_i} a_J^{-\alpha_i} +\sum_{JKL\beta_j\delta_k\xi_l\theta_m}\frac{\nu^{\alpha_i\beta_j\delta_k\xi_l}_{IJKL}\langle\theta_m|\xi_l\delta_k\rangle\langle-\alpha_i|\theta_m\beta_j\rangle}{\langle\alpha_i -\alpha_i|Ag\rangle} CDD_{JLK}^{-\alpha_i[\beta_j \theta_m[\xi_l\delta_k]]} $\\
 $\hat{A}^{\gamma_m}_{IJ} = CC^{\gamma_m[\alpha_i\beta_j]}_{IJ}$ &&&  $\hat{P}^{\gamma_m}_{IJ} =  \sum_{KL\delta_k\xi_l\alpha_i\beta_j}\frac{\nu^{\alpha_i\beta_j\delta_k\xi_l}_{IJKL}\langle-\gamma_m|\xi_l\delta_k\rangle\langle\gamma_m|\alpha_i\beta_j\rangle}{\langle\gamma_m -\gamma_m|Ag\rangle} DD^{-\gamma_m[\xi_l\delta_k]}_{LK}$\\
$\hat{B}^{\gamma_m}_{IJ} = CD^{\gamma_m[\alpha_i\beta_j]}_{IJ}$ &&&  $\hat{Q}^{\gamma_m}_{IJ} =  \sum_{KL\delta_k\xi_l\alpha_i\beta_j}\frac{(\nu^{\alpha_i\delta_k\beta_j\xi_l}_{IKJL}-\nu^{\alpha_i\delta_k\xi_l\beta_j}_{IKLJ})\langle-\gamma_m|\xi_l\delta_k\rangle\langle\gamma_m|\alpha_i\beta_j\rangle}{\langle\gamma_m -\gamma_m|Ag\rangle} CD^{-\gamma_m[\delta_k\xi_l]}_{KL}$\\
 \hline
  \end{tabular}
\end{table*}

A casual look at the redefinition of the operators in Table~\ref{tab:tops} suggests that they provide no computational benefit because although there are fewer tensor operators, 
each tensor operator hides within it a multiplet of individual operators. Further, even though all the formulae in Equation~\ref{eq:topgen} are given 
in terms of these tensor operators, the definition of the outer product $\otimes^\gamma$  in Equation~\ref{eq:defop} ensures that all the individual operators are involved in its evaluation. 
However, this is where the ``magic'' of spin algebra and the Wigner-Eckart theorem enters. We have already seen that due to the Wigner-Eckart theorem we only ever need store the reduced 
matrix elements of tensor operators. More importantly, during the blocking step the reduced matrix elements of the tensor operators on the system block, constructed using the formulae in Equation~\ref{eq:topgen}, can always be formed with knowledge only of the reduced matrix elements of the tensor operators on the left and dot block, by making use of Equation~\ref{eq:9j}. Thus, by working 
with the tensor operators in Table~\ref{tab:tops} and utilizing various predefined coefficients such as the Clebsch-Gordan coefficients, Racah's coefficient and 9-j coefficient of the 
non-Abelian group of interest, very substantial computational savings are possible.
\begin{align}
 \langle\phi_1^{\alpha_1}&\otimes^\alpha \phi_2^{\alpha_2}||O_1^\gamma \otimes^\xi O_2^\delta||\psi_1^{\beta_1}\otimes^\beta\psi_2^{\beta_2}\rangle = \nonumber\\
&\sum_{\psi_1\psi_2\phi_1\phi_2} \left[\begin{matrix}
         \beta_1 & \beta_2 &\beta\\
         \gamma & \delta & \xi\\
         \alpha_1 & \alpha_2& \alpha
        \end{matrix}\right] \langle\phi_1^{\alpha_1}||O_1^\gamma ||\psi_1^{\beta_1}\rangle \langle\phi_2^{\alpha_2}|| O_2^\delta||\psi_2^{\beta_2}\rangle\label{eq:9j} 
\end{align}

We  end this section on blocking with a few words about building multiplet states in a block. The full Fock space of an individual spatial orbital $I$ of irreducible representation $\alpha$ contains $2^{n_\alpha}$ many body states. These states do not form a basis for an irreducible representation of the symmetry group and have to be transformed using a unitary operator to form a set of multiplet states. This can be done in many ways including, by using projection operators\cite{hamermesh}, or by just acting the elements of a tensor operator on the vacuum state. The resulting number of multiplets obtained is less than $2^{n_\alpha}$. For example, in SU(2) symmetry each spatial orbital gives three set of multiplets, and in $SU(2)\times D_{\infty h}$ symmetry, one can get 
up to 7 sets of multiplets from 16 individual states.

\vspace{2 mm}

\noindent\textit{Wavefunction solution}: \\
The Hilbert space used to express the DMRG wavefunction of symmetry $\gamma$ is formed by the tensor product of the renormalized Fock space of the system and the 
environment block as shown in Equation~\ref{eq:wv}. In this space the Hamiltonian is diagonalized using the Davidson algorithm, which requires the ability to perform 
Hamiltonian-wavefunction multiplication ($|v\rangle = H|c\rangle$). This operation can be performed just in terms of multiplets as shown in Equation~\ref{eq:Hpsi}. Note that to reduce the scaling of the algorithm from $M^4$ to $M^3$ the explicit matrix representation of the Hamiltonian is never generated.
\begin{align}
& ||C^\gamma\rangle = \mathbf{C}^\gamma_{\phi_1^\alpha\phi_2^\beta} ||\phi_1^\alpha\rangle\otimes^\gamma||\phi_2^\beta\rangle \label{eq:wv}\\
& \mathbf{V}^\alpha_{\phi_1^\alpha\phi_2^\beta}   = \nonumber\\
&\sum_{\psi_1\psi_2\phi_1\phi_2} \left[\begin{matrix}
         \beta_1 & \beta_2 &\alpha\\
         \gamma & \delta & A_g\\
         \alpha_1 & \alpha_2& \alpha
        \end{matrix}\right] \langle\phi_1^{\alpha_1}||O_1^\gamma ||\psi_1^{\beta_1}\rangle \langle\phi_2^{\alpha_2}|| O_2^\delta||\psi_2^{\beta_2}\rangle \mathbf{C}^\alpha_{\psi_1^\alpha\psi_2^\beta}\label{eq:Hpsi} 
\end{align}
\vspace{2 mm}

\noindent\textit{Renormalization}: \\
During the renormalization step the reduced density matrix of the system is formed and its most significant eigenvalues are retained. When using non-Abelian symmetries in DMRG, instead of 
the usual reduced density matrix the ``quasi-density matrix'' which belongs to the fully symmetric irreducible representation is formed, as shown in Equation~\ref{eq:redden}. 
Diagonalizing the quasi-density matrix ensures that the multiplets belonging to different irreducible representations do not mix together during renormalization.
\begin{align}
 \Gamma^{A_g}_{\phi_1^\alpha\psi_1^\alpha} = \sum_{\phi_2^\beta} \mathbf{C}^\gamma_{\phi_1^\alpha\phi_2^\beta} \mathbf{C}^\gamma_{\psi_1^\alpha\phi_2^\beta} \label{eq:redden}
\end{align}
\vspace{2 mm}
\begin{table*}
\caption{DMRG energy (E+75.0 in E$_h$) of various states of the C$_2$ molecule calculated using the cc-pVQZ basis set with a frozen-core. The second column shows the ground state energies calculated to high accuracy (error $\approx$ 10 $\mu$E$_h$) with the number of renormalized (multiplet) states $M$, of up to 6000. The last 12 columns show the result of four separate 
DMRG state-averaged calculations, performed for three states each of the irreducible representations  $^1\Sigma^+_g$,  $^1\Sigma^+_u$, $^1\Sigma^-_g$ and $^1\Sigma^-_u$ with a maximum $M$ of 
4000. The difference between the second and the third column gives us an estimate of the error in our state-averaged calculations, which is less than 0.1 $m$E$_h$ along the entirety of the curve.}\label{tab:energies}
\begin{tabular}{lccccccccccccccccc}
\hline
\hline
r&$1^1\Sigma^+_g$&&\multicolumn{3}{c}{$^1\Sigma^+_g$}&&\multicolumn{3}{c}{$^1\Sigma^+_u$}&&\multicolumn{3}{c}{$^1\Sigma^-_g$}&&\multicolumn{3}{c}{$^1\Sigma^-_u$}\\
\cline{1-2}\cline{4-6}\cline{8-10}\cline{12-14}\cline{16-18}
1.1	& -0.76125	&& -0.76124	& -0.62183	& -0.50228	&& -0.56509	& -0.29484	& -0.22278	&& -0.28627	& -0.24395	& -0.21448	&& -0.41325	& -0.30927	& -0.19275\\
1.2	& -0.79924	&& -0.79920	& -0.69459	& -0.54490	&& -0.60050	& -0.35581	& -0.29406	&& -0.35914	& -0.30343	& -0.29363	&& -0.50018	& -0.44533	& -0.27846\\
1.24253	& -0.80269	&& -0.80264	& -0.71208	& -0.54953	&& -0.60338	& -0.36859	& -0.32773	&& -0.37614	& -0.32626	& -0.30224	&& -0.52284	& -0.48520	& -0.30105\\
1.3	& -0.79937	&& -0.79933	& -0.72633	& -0.54871	&& -0.59977	& -0.37819	& -0.36339	&& -0.39013	& -0.34767	& -0.30515	&& -0.54539	& -0.52476	& -0.32009\\
1.4	& -0.77970	&& -0.77965	& -0.73267	& -0.53776	&& -0.58049	& -0.41085	& -0.37409	&& -0.40043	& -0.36328	& -0.30354	&& -0.58136	& -0.55243	& -0.33551\\
1.6	& -0.72405	&& -0.72401	& -0.70487	& -0.51054	&& -0.52367	& -0.45514	& -0.34627	&& -0.40408	& -0.33810	& -0.29200	&& -0.61877	& -0.54743	& -0.35138\\
2	& -0.64560	&& -0.64552	& -0.61469	& -0.49290	&& -0.47373	& -0.42377	& -0.32265	&& -0.38251	& -0.31599	& -0.27675	&& -0.61374	& -0.50634	& -0.43199\\
\hline
\end{tabular}
\end{table*}

\noindent\textit{Carbon dimer}: \\
C$_2$ is an important intermediate in combustion processes and its electronic spectrum has been the subject of many experiments\cite{weltner, Martin1992, orden}. Also, 
due the  multi-reference nature of the ground and excited states of C$_2$ at stretched bond lengths, it has attracted theoretical interest as a benchmark system for methods such as 
coupled-cluster\cite{Watts1992, Sherrill2005,Lyakh2008, Datta2011}, full configuration interaction\cite{Abrams2004}, multireference perturbation theory\cite{Mahapatra2008,Jiang2011}, multireference configuration interaction\cite{Pradhan1994}, initiator full configuration interaction quantum Monte Carlo (i-FCIQMC)\cite{Booth2011, Cleland2012}, model space quantum monte carlo (MSQMC)\cite{tenno-msqmc, tenno-msqmc2}, correlation energy extrapolation by intrinsic scaling (CEEIS)\cite{Bytautas2005}, reduced-density-matrix theory\cite{Gidofalvi2005} and valence bond theory\cite{Shaik2012}. Here we apply our non-Abelian symmetry adapted DMRG to calculate 
12 low energy states - the 3 lowest energy levels each for irreducible representations $^1\Sigma^+_g$,  $^1\Sigma^+_u$, $^1\Sigma^-_g$ and $^1\Sigma^-_u$. Note the states of irreducible representation $^1\Sigma^-_g$ and $^1\Sigma^-_u$ cannot be easily targeted,  when only Abelian subgroups of $D_{\infty h}$ are used. Further the use of the full symmetry of the problem allows us to 
very efficiently obtain high accuracy, which we exploit to affordably compute not just the ground, but also the excited states of each of the irreducible representations.

\begin{figure}
\begin{center}
\includegraphics[width=0.4\textwidth]{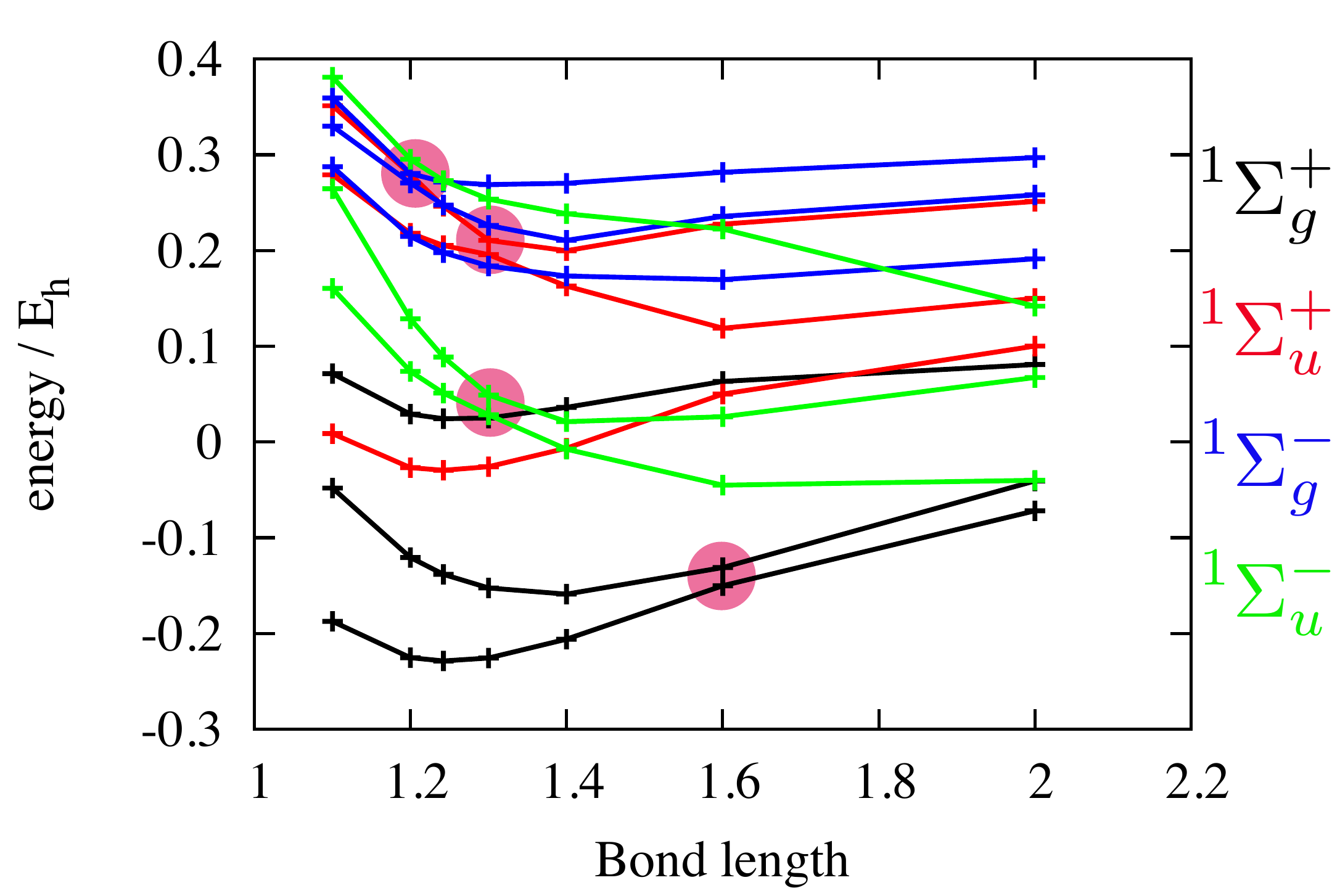}
\end{center}
\caption{The energies of various states of the C$_2$ molecule relative to complete dissociation, consisting of two carbon atoms in triplet states. The shaded circles show the regions of avoided crossing between states belonging to the same irreducible representation. As expected, we see that the energy of three singlet states: two $^1\Sigma^+_g$ and one $^1\Sigma^-_u$, approach the value of two carbon triplet atoms at stretched bonds.\label{fig:dmrgm2}}
\end{figure}

A ground state Hartree-Fock calculation was performed on the C$_2$ molecule with the cc-pVQZ\cite{dunningbasis} basis set using the \textsc{Molpro}\cite{werner2012} quantum chemistry package. The orbitals obtained from the Hartree-Fock calculation were first transformed to make them compatible with the Clebsch-Gordan coefficients for the $D_{\infty h}$ point group, as given 
in Altmann and Herzig\cite{altmann}. The core orbitals and electrons were frozen, but the remaining 8 electrons in 108 orbitals were fully correlated. First, only the ground state calculation was performed using DMRG; the resulting energy is tabulated in the second column of Table~\ref{tab:energies}. By fitting the DMRG energy versus the discarded weight to a quadratic curve the 
remaining energy error was estimated to be less than 10 $\mu$ E$_h$. The calculated DMRG energy is more accurate than the i-FCIQMC\cite{Cleland2012} and the CEEIS\cite{Bytautas2005} calculations 
and well within their reported error bars of 0.3 mE$_h$. Next, four separate DMRG state-averaged calculations were performed on three states each of irreducible representations  $^1\Sigma^+_g$,  $^1\Sigma^+_u$, $^1\Sigma^-_g$ and $^1\Sigma^-_u$ with a maximum $M$ (number of retained renormalized (multiplet) states) of 4000. The energies of all the states are shown in columns 3 to 14 of Table~\ref{tab:energies}. The difference between the energies of the first and the second column gives an estimate of error of our state-averaged calculations and is less than 0.1 mE$_h$ for all the geometries. These energies are plotted in Figure~\ref{fig:dmrgm2} and show several curve crossings and avoided curve crossings.

In summary, we have presented an efficient extension of our spin-adapted DMRG algorithm to efficiently utilize general non-Abelian point group symmetries. 
We used the resulting implementation  to calculate the ground state 
potential energy curve of the C$_2$ molecule with a cc-pVQZ basis set (and frozen core) to an unprecedented (near-exact) accuracy of 
10 $\mu$E$_h$. Further, several excited state potential energy curves were also computed to very high accuracy to demonstrate the large efficiency gains in this new formulation. These calculations
may for practical chemical purposes be considered exact within the given basis set. Finally, as we have recently demonstrated elsewhere\cite{sharmabe2}, the remaining
basis set error can be eliminated by combining efficient DMRG algorithms for large bases as described in this work, with the transcorrelated Hamiltonian approach 
of Shiozaki and Yanai\cite{yanai2012canonical}, thus enabling solutions of the molecular Sch\"odinger equation to spectroscopic accuracy.


\begin{acknowledgements}
This work was supported by the US National Science Foundation (NSF) through Grant No. 
NSF-CHE-1265277. Additional support for software development was provided
through Grant NSF-OCI-1265278. S. S. would also like to thank Garnet Chan for many helpful discussions.
\end{acknowledgements}


\end{document}